%% file: main.tex
\documentclass[
  prl,
  twocolumn,
  superscriptaddress,
  citeautoscript,
  showpacs,
  amsart,
  longbibliography
]{revtex4-2}

\usepackage{amsmath,bm,amsfonts}
\usepackage{physics,mathtools}
\usepackage{graphicx}
\usepackage[normalem]{ulem}
\usepackage{titlesec}
\usepackage[colorlinks=true,linkcolor=blue,citecolor=blue]{hyperref}
\usepackage[dvipsnames]{xcolor}
\usepackage[english]{babel}
\usepackage{amsthm} %for proof
\newtheorem{proposition}{Proposition}
\newtheorem{theorem}{Theorem}
\newtheorem{lemma}[theorem]{Lemma}

\begin{document}

\title{Measurement-induced phase transitions in systems with diffusive dynamics}

\author{Hyunsoo Ha}
\affiliation{Department of Physics, Princeton University, Princeton, NJ 08544, USA}
\author{Akshat Pandey}
\affiliation{Department of Physics, Stanford University, Stanford, CA 94305, USA}
\author{Sarang Gopalakrishnan}
\affiliation{ Department of Electrical and Computer Engineering, Princeton University, Princeton, NJ 08544, USA}
\author{David A. Huse}
\affiliation{Department of Physics, Princeton University, Princeton, NJ 08544, USA}

\date{\today}

\begin{abstract}
The competition between scrambling and projective measurements can lead to measurement-induced entanglement phase transitions (MIPT).
In this work, we show that the universality class of the MIPT can be drastically altered when the system has a diffusing conserved density.
As a numerical tractable model of this, we study a 1+1d random Clifford circuit locally monitored by classically diffusing particles (``measurers'').  The resulting diffusive correlations in the measurement density are a relevant perturbation to the usual space-time random MIPT critical point, producing a new universality class for this phase transition.
We find ``Griffiths-like'' effects due to rare space-time regions where, e.g., the diffusive measurers have a low or high density, but these are considerably weaker than the Griffiths effects that occur with quenched randomness that produce rare spatial regions with infinite lifetime.
\end{abstract}

\maketitle

%%%%%%%%%%%%%%%%%%%%%%%%%%%%%%%%%%%%%%%%%%%%%%%%%%%%%%%%%%%%%%%%%

Measurement-induced phase transitions (MIPTs) are a recently discovered class of phase transitions in quantum dynamical systems subject to repeated measurements. MIPTs are transitions in the entanglement structure of typical individual \emph{quantum trajectories}---i.e., in the state of the system conditional on a set of measurement outcomes~\cite{li_fisher_quantumzeno, skinner_nahum_prx2019, Li_Chen_Fisher}. They separate a ``mixed'' or ``volume-law'' phase in which the dynamics generate volume-law entanglement that is able to preserve some quantum information about the initial state, from a ``pure'' or ``area-law'' phase in which measurements rapidly extract all quantum information 
and prevent the buildup of entanglement~\cite{PhysRevLett.125.030505, Gullans_Huse_purification}. The MIPT in spatially local one-dimensional random quantum circuits, with local unitary gates and single-site projective measurements, has been the best-studied example~\cite{vasseur+ludwig_MIPT+RTN, bao_altman_statmech+MIPT, Zabalo_Pixley_diagnostics, li_fisher_CFT+MIPT, zabalo_pixley_multifractal+MIPT, potter2022entanglement, fisher2023random}; however, MIPTs have also been explored in other geometries~\cite{Turkeshi_2020_2+1D,PhysRevB.102.064202, vijay2020measurement, PhysRevB.104.155111, PRXQuantum.2.010352,Turkeshi_2022_3+1D, PRXQuantum.4.030333}. Experimental studies of MIPTs remain challenging because studying individual trajectories requires procedures (such as post-selection or classical simulations) for which the resource requirements grow exponentially with system size; nevertheless, signatures of MIPTs have been seen in recent experiments with trapped ions and superconducting qubits~\cite{noel2022measurement, koh2023measurement, hoke2023quantum, kamakari2024experimental}, and recent theoretical proposals indicate how MIPTs may be realized in larger-scale experiments~\cite{PhysRevLett.130.220404, PhysRevLett.129.200602, garratt2023probing}.

The critical properties of MIPTs depend on the spatial dimension and range of the gates~\cite{PhysRevB.104.155111, Minato_Saito_longrangeMIPT1, Block_Yao_longrangeMIPT2, Muller_Buchhold_longrangeMIPT3}. In addition, typical trajectories are inherently random in space and time---since the measurement outcomes are random---and the critical properties are sensitive to the correlations in this randomness. The ``standard'' MIPT occurs in one-dimensional circuits where the unitary gates and measurement locations are chosen at random with no spacetime correlations, and the measurement outcomes are set by the Born rule. A number of circuit-averaged observables at this standard MIPT seem to be governed by a nonunitary conformal field theory~\cite{li_fisher_CFT+MIPT}, which, in the limit of infinite on-site Hilbert space dimension, can be identified as percolation. When the measurement locations or the gate parameters have strong enough long-range correlations, the universality class changes. For example, if there is quenched spatial randomness in the measurement probabilities, the transition is instead governed by an infinite-randomness fixed point with infinitely anisotropic spacetime scaling~\cite{Zabalo_Pixley_infiniterandomnessMIPT, shkolnik_gazit_quasiperiodicMIPT}.

% In the standard MIPT, the dynamics is not subject to any conservation laws. Even in the absence of measurements, conservation laws constrain entanglement growth in quantum circuits \cite{Rakovsky2021diffusive_renyi}. 
% In this paper we explore how the presence of a diffusing conserved charge (conserved also by the measurements) affects the MIPT in 1+1 dimensions.  Generically, the measurement rate at the transition will depend linearly on small changes in the overall charge density~\cite{chakraborty_Pixley_U1MIPT}.  When this is true, as we argue below, the coupling to the diffusing charge is relevant at the critical point, and thus changes the universality class of this phase transition~\footnote{At special points where the critical measurement rate does not depend linearly on the density, the coupling to diffusion may be irrelevant. We will not consider this case in the present work.}. Note that, in addition to the MIPT (which is an \emph{entanglement} transition), circuits with a conserved charge also exhibit a ``charge-sharpening'' phase transition~\cite{Agrawal_Vasseur_U1MIPT, Barratt_Potter_U1MIPTfieldtheory, PhysRevLett.129.200602, Majidy_Halpern_SU2MIPT}. The MIPT takes place in the charge-sharp phase, in which measurements have collapsed the charge distribution, so an individual quantum trajectory has a well-defined spacetime charge density profile at large length scales.

In the standard MIPT, the dynamics is not subject to any conservation laws. Even in the absence of measurements, conservation laws constrain entanglement growth in quantum circuits \cite{Rakovsky2021diffusive_renyi}. 
In this paper we explore how the presence of a diffusing conserved charge (conserved also by the measurements) affects the MIPT in 1+1 dimensions.  
Our analysis also extends to the $U(1)$ symmetric monitored circuits, where in addition to the MIPT (which is an \emph{entanglement} transition), circuits with a conserved charge also exhibit a ``charge-sharpening'' phase transition~\cite{Agrawal_Vasseur_U1MIPT, Barratt_Potter_U1MIPTfieldtheory, PhysRevLett.129.200602, Majidy_Halpern_SU2MIPT}. Note that the MIPT takes place in the charge-sharp phase, where the measurements are frequent enough to collapse and localize the charge distribution, so an individual quantum trajectory has a well-defined spacetime charge density profile that random-walks at large length scales.  Generically, the critical {measurement rate at the entanglement transition depends linearly on small changes in the overall charge density~\cite{chakraborty_Pixley_U1MIPT}. When this is true, as we argue below, the coupling to the diffusing charge is relevant at the critical point, and thus changes the universality class of this phase transition~\footnote{At special points where the critical measurement rate does not depend linearly on the density, the coupling to diffusion may be irrelevant.  We will not consider this case in the present work.}.
 
The MIPT can be studied efficiently for large systems composed of Clifford gates and Pauli measurements \cite{Li_Chen_Fisher,Zabalo_Pixley_diagnostics,Gullans_Huse_scalableprobe,Gullans_Huse_purification,li_fisher_CFT+MIPT,zabalo_pixley_multifractal+MIPT}.  Unfortunately, if one restricts such a random Clifford circuit to have a conserved $U(1)$ charge (e.g. $\hat{Z}_i$ summed over all qubits $i$) this eliminates the volume-law entangled phase and thus also the phase transition that we wish to study (see \cite{supp_mat} for detailed discussion). Thus we introduce a new type of model in order to efficiently numerically study this transition.
In this model, a diffusing conserved {\it classical} charge is separate from the quantum Clifford random circuit, the circuit alone has no conservation laws, and the measurement rate at a given space-time point is set by the (classical) charge at that point.  This allows us to simulate large systems and thus investigate the resulting new universality class of the MIPT.  In the absence of the conserved charge, the universality class of the transition does differ in some details between Clifford and non-Clifford (e.g., Haar-random) systems, although in many respects they are qualitatively similar \cite{Zabalo_Pixley_diagnostics,zabalo_pixley_multifractal+MIPT}.  With the conserved charge, we can again expect differences between non-Clifford and Clifford universality classes; here, we study the latter because it is much more accessible numerically by using the new type of model that we introduce. Our main results are to locate and characterize the critical point of the MIPT in this model. We also provide numerical evidence and analytical arguments to show that diffusive correlations affect the properties of the phases on either side of the MIPT.

Our model has classical particles diffusing via the symmetric simple exclusion process (SSEP) \cite{Liggett_SSEP,kipnis_SSEP,eyink_SSEP} on the same one-dimensional lattice that the qubits occupy.  The particle density is equal to one particle per two qubits.  These classical particles are the ``measurers'': at the time of measurements, each qubit that is at a site occupied by a particle is measured with probability $p$; see Fig.~\ref{fig:model}. 
This imparts diffusive correlations in space-time to the locations of the measurements.
In all data presented here, the measurers have diffusivity $D=1$, and we use periodic boundary conditions except where otherwise stated. Our systems are of length $N$ qubits.

\begin{figure}
    \centering
\includegraphics[width=0.5\textwidth]{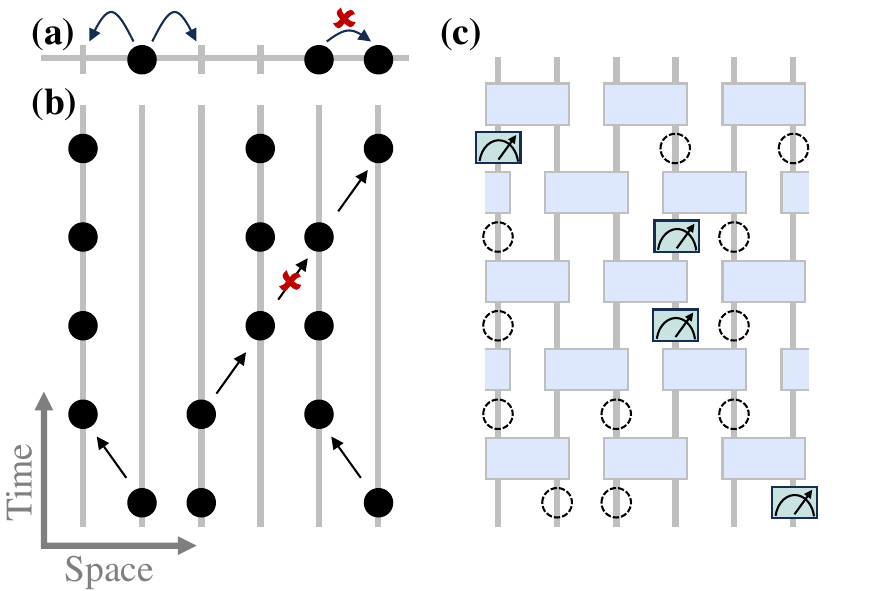}
    \caption{Random Clifford circuit monitored by diffusing classical particles. (a) Illustration of classical SSEP rules governing measurer dynamics, where hopping to nearest-neighbor sites occurs as a continuous-time Poisson process with a specified rate and is forbidden if the adjacent site is already occupied. (b) 
    An example of measurer trajectories.  (c) A circuit with measurer locations as in (b). 
    Two-qubit random Clifford gates are applied in a brickwork manner. At each time between layers of gates, each measurer makes a $\hat{Z}$ projective measurement of the local qubit with probability $p$.} 
  \label{fig:model}
\end{figure}

\emph{\textbf{Generalized Harris argument---}} We begin by giving a generalized Harris argument~\cite{Vojta_generalized_Harris, ando_huse} for the relevance of diffusive correlations in the measurement rate.
The standard MIPT (with uncorrelated measurements) has correlation length exponent $\nu$  and dynamical exponent 
$z=1$, so for $p$ near the critical point $p_c$ the correlation region has space-time volume $\sim \xi^{(d+1)}(p) \sim |p-p_c|^{-(d+1)\nu}$ for $d+1$ dimensions. 
Space-time correlations in the measurements will be relevant if they cause the measurement rate averaged over a space-time correlation volume to vary (between different correlation volumes of the same system) by an amount $\delta p$ that vanishes more slowly than $|p-p_c|\sim \xi^{-1/\nu}$ as $p\rightarrow p_c$.

If diffusive measurers are added perturbatively to the standard uncorrelated model, the number of measurers in a correlation volume does not change significantly within one correlation time; the measurer number is strongly correlated along the time direction. The variation in the number of measurers is proportional to the square root of the spatial volume, resulting in a variation of measurer density within a correlation volume proportional to $\xi^{-d/2}$. Therefore, $\delta p \sim \xi^{-d/2}(p)$, which is relevant when $\nu<2/d$, as in the usual Harris argument \cite{Harris_criterion_1974}.  
In particular, for the 1+1-dimensional case that we consider here, the uncorrelated critical point has $\nu\cong 1.3<2/d=2$. Therefore, diffusive correlations are indeed a relevant perturbation for the uncorrelated MIPT. Note that this relevance criterion does not change as long as the measurers move slower than ballistically (so that the number of measurers within a correlation volume does not change significantly in one correlation time in the $p\rightarrow p_c$ limit).

%%%%%%%%%%%%%%%%%%%%%%%%%%%%%%%%%%%%%%%%%%%%%%%%%%%%%%%%%%%%%%%%%%%%%%%%%%%%%%
\emph{\textbf{Critical point: numerical results---}} 
The generalized Harris argument establishes that diffusive correlations drive the system away from the spacetime-uncorrelated critical point. To study the new critical point that \emph{does} occur, we turn to numerical simulations of these Clifford circuits.  We begin by estimating the critical measurement rate $p_c$ based on the peak of the antipodal mutual information (AMI) defined as $S_A+S_B-S_{AB}$, where $A$ and $B$ are subsystems of length $N/8$ qubits located at opposite ends of the system [inset of Fig.~\ref{fig:critical point}(a)]. Here, $S_{A(B)}$ is the bipartite entanglement entropy of subsystem $A(B)$.
The AMI vanishes as $N \to \infty$ in both the volume-law and area-law phases and is expected to exhibit a peak at $p_c$ under fairly general conditions~\cite{Li_Chen_Fisher}.
As shown in Fig.~\ref{fig:critical point}(a), the peak in the AMI for accessible sizes suggests 
$p_c\cong 0.33$. Interpreted conservatively, our data on the AMI constrains $p_c$ to lie in the interval $(0.31, 0.35)$: for $p$ within this interval, the AMI grows with $L$ for all accessible sizes, so any value in this range could be the critical point. 
The estimated $p_c$ for this model is approximately twice that in the standard random-Clifford MIPT. This factor of roughly two is plausible because the number of measurers is $N_p=N/2$, so the average measurement rate over all spacetime is $p/2$. 
In \cite{supp_mat}, we attempt to extract the correlation-length exponent $\nu$ at this critical point by collapsing the AMI as well as a different observable, the tripartite mutual information. This gives us a rough estimate of $\nu = 1.6(5)$. The relatively poor quality of the collapse indicates that despite the large system sizes we are considering, finite-size effects are still severe.

To explore dynamical scaling at the critical point, we turn to the purification dynamics of an initially maximally mixed state for measurement rates $p$ around this critical window. We specifically investigate the purification time $\tau_P$ when the maximally mixed state purifies to a pure state. At the critical point, we expect that the timescale for purifying a system of size $N$ scales as $\tau_P(N) \sim N^z$, where $z$ is the dynamical critical exponent. For the standard MIPT, $z = 1$, and for the case of quenched randomness, $z = \infty$~\cite{Zabalo_Pixley_infiniterandomnessMIPT}. In the diffusive case, our numerical results [Fig.~\ref{fig:critical point}(b)] clearly indicate that $z \geq 2$. Assuming that $N = 256$ is larger than any relevant microscopic scale, the slope $d(\log \tau_P)/d(\log N)$ at $p = 0.35$ lower bounds the value of $z$. This lower bound is close to the diffusive value $z = 2$, which, as we will discuss, is the largest value that is theoretically well-motivated. If we instead focus on the apparent $p_c$---given by the peak in the AMI---we find a slope that is clearly drifting to larger values with increasing $N$. This would imply $z > 3.5$ and would be most consistent with a formally infinite $z$, with power-law dynamical scaling replaced by some form of activated scaling, as at the infinite-randomness critical point. Indeed, as shown in the inset to Fig.~\ref{fig:critical point}(b), an activated scaling ansatz with scaling $\log \tau_P \sim N^{\psi}$ with $\psi \cong 0.44$ is consistent with the purification data.

 \begin{figure}
    \centering
    \includegraphics[width=0.5\textwidth]{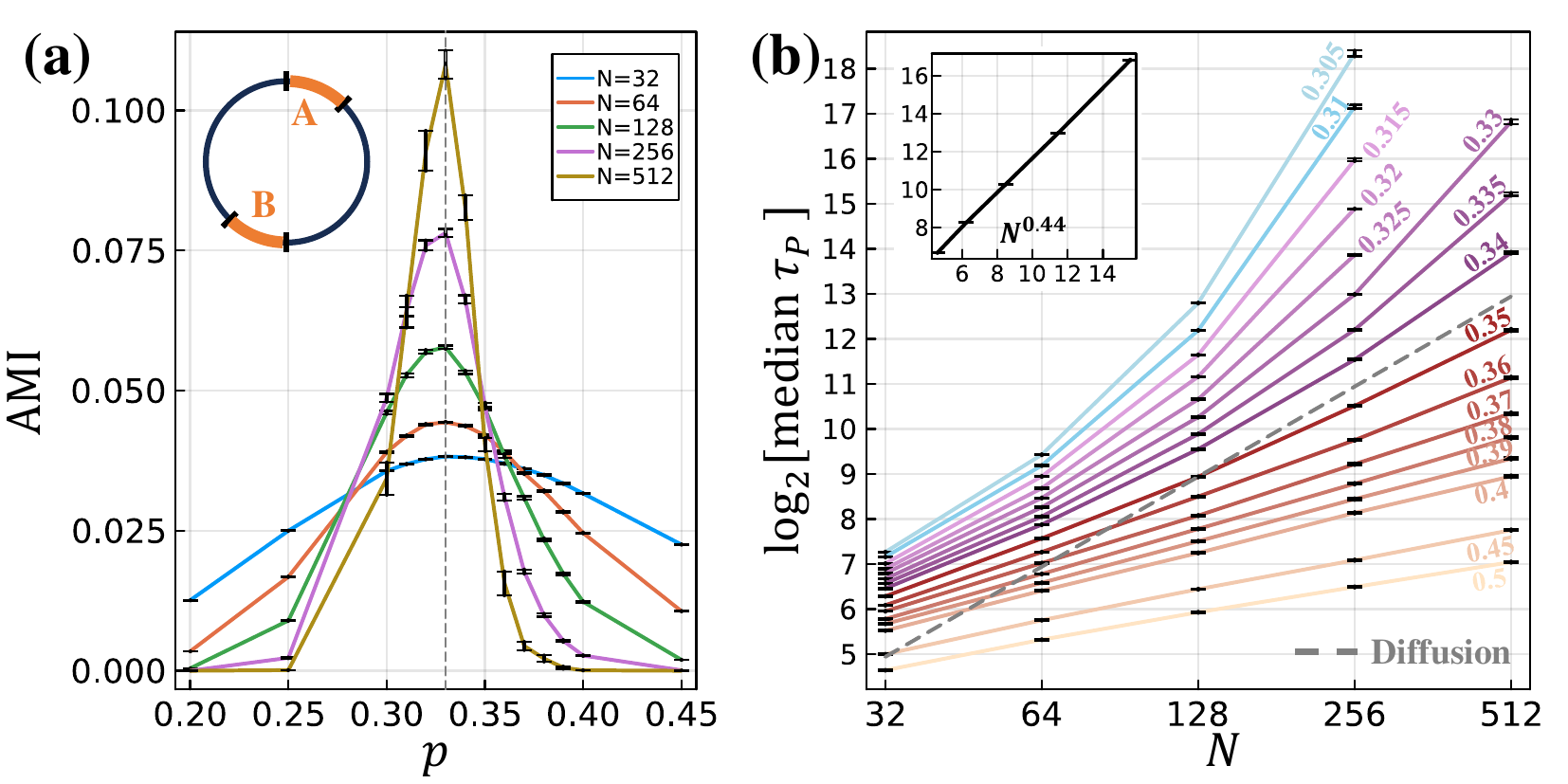}
    \caption{Entanglement and purification dynamics at and near the critical point. (a) Antipodal mutual information in the long-time steady state distribution of pure states shows that the critical point $p_c$ is located near $p=0.33$. (b) Purification time $\tau_P$. The median over samples is presented as a representative of a typical sample. 
    The gray dotted line is the relaxation time scale for the diffusive measurers with $z=2$. The inset shows the apparent activated dynamical scaling at $p=0.33$, with the horizontal axis re-scaled to $N^{0.44}$. 
    }
  \label{fig:critical point}
\end{figure}

Thus, our numerical results are consistent with at least two possibilities for the critical point. One possible scenario, suggested by a straightforward analysis of the numerics, gives $p_c \approx 0.33$ and $z > 3.5$. Although consistent with the numerics, this scenario seems implausible, by the following logic: Imagine running a renormalization-group scheme that gives an exponent $z > 2$. After many steps of this scheme, the system will have $O(1)$ remaining degrees of freedom per $\ell \times \ell^z$ patch for some large $\ell$. The remaining disorder at this scale is spatially uncorrelated, so if the renormalized description of the circuit resembles the original microscopic description, the flow beyond this scale will be governed by the standard MIPT and $z$ will decrease. We emphasize that this argument makes some nontrivial assumptions about the nature of the flow, and does not rule out the $z > 2$ scenario. A simpler possibility, however, is that the AMI is showing strong finite-size effects and the true critical point lies near the upper end of the allowed range $(0.31, 0.35)$ and has $z = 2$, so the apparent values of $z > 2$ are actually in the volume-law phase.  We present some tentative numerical support for this possibility in \cite{supp_mat}.

%%%%%%%%%%%%%%%%%%%%%%%%%%%%%%%%%%%%%%%%%%%%%%%%%%%%%%%%%%%%%%%%%%%%%%%%%%%%%%
%%%%%%%%%%%%%%%%%%%%%%%%%%%%%%%%%%%%%%%%%%%%%%%%%%%%%%%%%%%%%%%%%%%%%%%%%%%%%%
\emph{\textbf{Rare region effects---}} 
We now turn to the phases on either side of this new MIPT. As in the infinite-randomness case, we expect that some properties of these phases are dominated by rare spacetime regions that are locally ``in the other phase''. To influence dynamics on a timescale $t$, such a rare configuration must persist for time $t$. In the standard MIPT, the measurement probabilities are drawn independently at each time, so this probability decays exponentially in time. However, in the diffusive case, a rare region of size $\ell \agt \sqrt{Dt}$ typically persists for time $t$. The probability ``cost'' of such a region is stretched-exponential in $t$, so rare-region effects can dominate late-time dynamics of certain quantities whose typical behavior is exponential decay. The quantities that show these effects are different in the two phases:

In the volume-law phase, we find that regions with anomalously high density of measurers behave as bottlenecks that suppress \textit{entanglement growth} due to the resulting high density of measurements.
We prepare a random pure product initial state, evolve with open boundary conditions until time $t$, bipartition the system with a single cut at position $x$ near the midpoint ($x=N/2$) and obtain the entanglement entropy that we label $S_{N/2}(t)$. We define the total probability $P_{S_{N/2}}^<(t)$ at time $t$ of having $S_{N/2}(t) < \bar{S}_{N/2}(t)/2$: 
the probability of the entanglement entropy for a single instance being suppressed by at least a factor of two from its mean over many instances at time $t$. 

In the usual MIPT, the most probable way to produce such a large suppression is due to a space-time path of time duration $\sim t$ that has substantially suppressed entanglement growth along the entire path. Therefore one expects $-\log P_{S_{N/2}}^<\sim t$.   In our diffusive circuit, the tail is due to the rare regions of spatial width $O(\sqrt{Dt})$ such that they can survive until time $t$. Therefore, we expect $-\log P_{S_{N/2}}^< \sim \sqrt{Dt}$.  With time-independent quenched randomness, the rare region is a bottleneck of the entanglement growth with a rate exponentially decaying with its length. In this case the most probable way to suppress the entanglement is due to a nearby bottleneck that is of spatial width $O(\log t)$, leading to the Griffiths power law: $-\log P_{S_{N/2}}^< \sim \log t$ \cite{Zabalo_Pixley_infiniterandomnessMIPT}.  We numerically verified the results are consistent with expectations for all three models (Fig.~\ref{fig:rare_region_volume}).
This probability also oscillates in time as it decays because entanglement in Clifford circuits is integer-valued \cite{supp_mat}.
\begin{figure}
    \centering
    \includegraphics[width=0.5\textwidth]{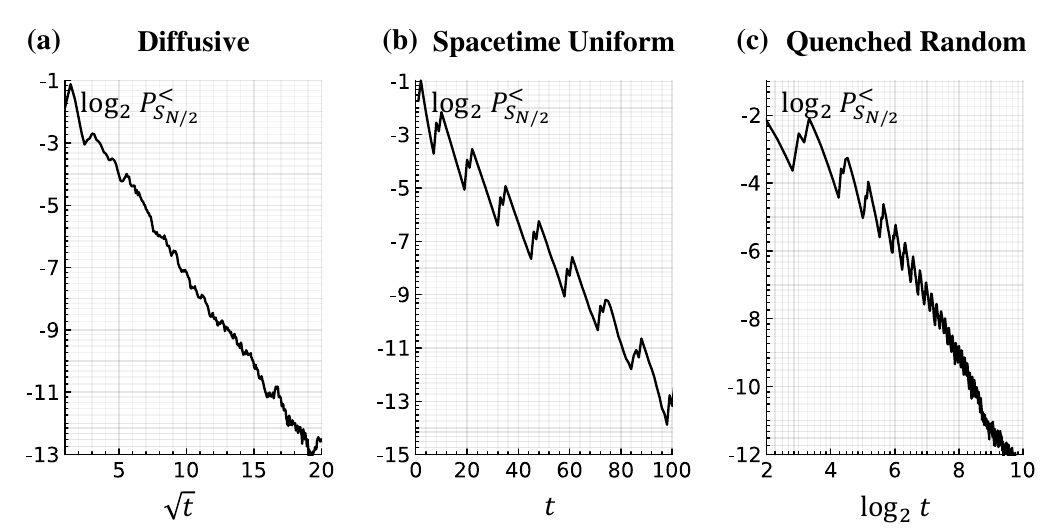}
    \caption{Rare region effects in the volume-law phase in models with (a) diffusing measurers, (b) space-time uncorrelated measurements, and (c) immobile measurers having quenched randomness.  Note, the horizontal axis is different in each case.  The presented measurement rates are $p\cong p_c/2$, thus well into the volume-law phase.  The system size is $N=1024$, and cuts are sampled within a distance of 64 sites of the center of the chain.} 
  \label{fig:rare_region_volume}
\end{figure}

In the area-law phase, rare regions with low measurement density slow down
\textit{purification} from the maximally mixed initial state. We focus on two quantities: the mean total entropy $S(t)$ (entropy of the mixed state averaged over samples) and the tail distribution of the purification time $\tau_P$, denoted as $P_{\tau_P}^>(t)\equiv P(\tau_P \geq t)$. These two quantities behave similarly at very late times when $S(t)<1$, as both diagnose the last bit of entropy to be purified.

In the intermediate time regime of $1\ll Dt<O(N^2)$, the entropy remains in rare regions with a lower density of measurers that were already present at time zero.  For these to remain until time $t$ they must be of spatial length $\ell \gtrsim \sqrt{Dt}$. The probability of having a rare region decays exponentially in its length $\ell$, and once the probability is rare, there are approximately $N/\ell$ independent locations where such rare regions may occur.  Therefore, we expect the following intermediate-time scaling:
\begin{align}
\log(\frac{S(t)}{N/\sqrt{Dt}}) \sim -\sqrt{Dt},\quad\quad  \log(\frac{P_{\tau_P}^>(t)}{N/\sqrt{Dt}}) \sim -\sqrt{Dt}~,
\label{eqn:rare_region_area_early}
\end{align}
which are clearly observed in Fig.~\ref{fig:rare_region_area}(a,b).
The typical purification time diverges with system size as $\sim (\log N)^2/D$:  The largest rare region length is typically $O(\log N)$, and the typical purification time is determined by the diffusive lifetime ($z=2$) of this rare region.

\begin{figure}
    \centering
    \includegraphics[width=0.5\textwidth]{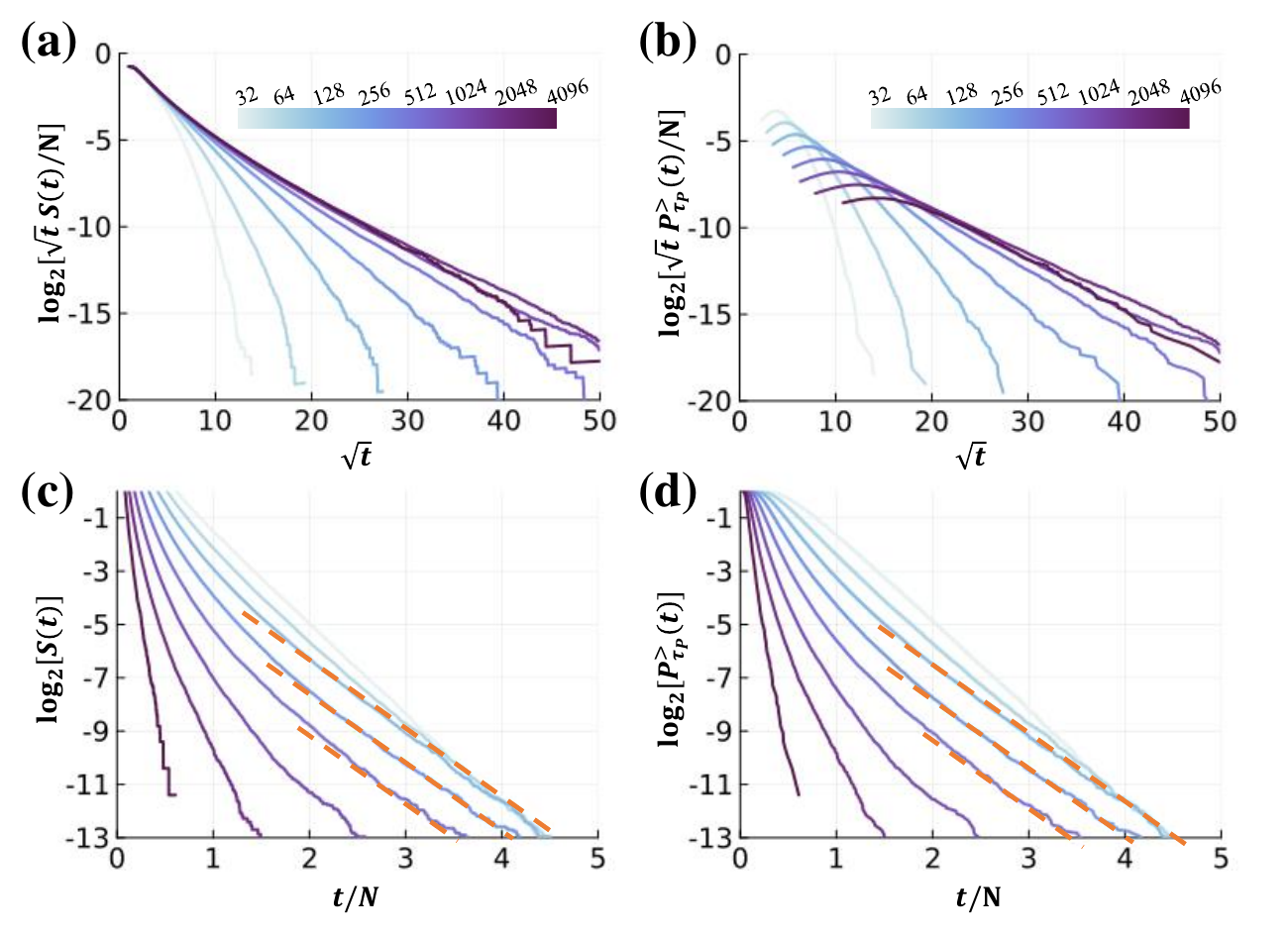}
    \caption{ Rare region effects in the area-law phase. (a,b) Numerical demonstrations of Eqs.~(\ref{eqn:rare_region_area_early}) at intermediate times, where the longest-living rare region determines the decay of the total entropy and the tail of the purification time distribution. (c,d) 
    At later times $S(t)$ and $P^>_{\tau_P}(t)$ decay exponentially with $t/N$. Color scale indicates $N$.}  \label{fig:rare_region_area}
\end{figure}

Beyond the diffusive time scale $t\gg O(N^2/D)$, almost all the rare samples in the area-law phase that have not fully purified have entropy one bit.  This logical operator has survived in spite of the strong measurements due to a sequence of rare spacetime regions.  The most probable way for a logical to survive for such long times is for there to be a rare region of size of order $N$: the density of measurers is low over some fraction of order one half of the full system.  Such a pattern occurs with a probability that is exponentially small in $N$.  This rare pattern would normally disappear after a time $\sim N^2/D$, so for it to persist even longer to time $t\gg N^2/D$ means it did not decay away of order $Dt/N^2$ times.  Thus the probability of the system not purifying to these late times is  
\begin{align}
\log P_{\tau_P}^>(t) \cong \log S(t) \sim (-N) \times \frac{Dt}{N^2} = -\frac{Dt}{N}.
\end{align}
This is consistent with Fig.~\ref{fig:rare_region_area}(c,d), where for large $t/N$ both $S(t)$ and $P_{\tau_P}^>(t)$ decay exponentially with $t/N$ (orange dotted lines are parallel).

\emph{\textbf{Conclusion}}--- We have shown that diffusive correlations due to a conservation law change the universality class of the MIPT, as well as the long-time asymptotics of the phases on either side of it---specifically, entanglement dynamics in the volume-law phase and purification dynamics in the area-law phase. We have provided analytic arguments to show that the universality class of the transition must change, and to estimate the ``Griffiths-like'' effects from diffusive correlations in the phases. To isolate the effect of diffusive correlations while still reaching large system sizes, we numerically explored a model in which the diffusive correlations are put in by hand. However, our analytic arguments do not rely on the specifics of this model, and we expect that our main conclusions are general, although critical exponents might differ between Clifford and generic circuits.

If $z> 2$, the correlation length exponent has a weaker lower bound, as the single correlation volume lasts for $\xi^{z-2}$ correlation times of the measurers on scale $\xi$.
Hence, the rms variation in the measurement rate between such space-time correlation volumes is $\delta p \sim \xi^{(2-d-z)/2}$, matching to the $\sim \xi^{d/2}$ for $z\leq 2$. This gives a generalized Chayes inequality $\nu\geq 2/(z+d-2) = 2/(z-1)$ for $1+1$ dimension and $z\geq 2$.
Therefore, if the system is behaving on some scale with apparent $z>2$, the apparent $\nu$ we extract may only obey this generalized inequality and ``explain'' our apparent $\nu<2$ (although it has a large uncertainty).

The critical point we have explored is one of a large family of new critical points that arise naturally in the dynamics of quantum and/or classical information. The sharpening transition is another example in this vein; in one dimension disorder is irrelevant at this transition, but in higher dimensions (or for symmetry groups other than $U(1)$) it might be a relevant perturbation. The dynamics of out-of-time-order correlators in classical diffusive spin chains is another closely related example~\cite{ando_huse}. It was also recently shown that the coupling between quantum and classical circuits leads to nontrivial entanglement growth \cite{klocke_moore_buchhold2024critical}. These critical points involve an information-theoretic quantity, such as entanglement, charge fluctuations, or an out-of-time-order correlator, coupled {nonreciprocally} to one or more hydrodynamic modes. Developing a framework for describing these new phase transitions is an important topic for future work.

\acknowledgments{H.H. thanks Grace Sommers and Su-un Lee for advice on numerics,  Zack Weinstein, Ahana Chakraborty, Vir Bulchandani, and Rhine Samajdar for fruitful discussions. A.P. is grateful to Aditya Cowsik and Nicholas O'Dea for discussions. S.G. and D.A.H. thank Jed Pixley, Drew Potter, and Romain Vasseur for helpful discussions and collaborations on related topics. D.A.H. was supported in part by NSF QLCI grant OMA-2120757. This material is based upon work supported by the U.S. Department of Energy, Office of Science, National Quantum Information Science Research Centers, Co-design Center for Quantum Advantage (C2QA) under contract number DE-SC0012704 (theoretical analysis performed by S.G.). Simulations presented in this work were performed on computational resources managed and supported by Princeton Research Computing. The open-source \texttt{QuantumClifford.jl} package was used for Clifford simulations \cite{quantumclifford}.}

\bibliographystyle{apsrev4-2} 
\bibliography{reference.bib}

\include{supp}

\end{document}

%% file: supp.tex
\newpage
\onecolumngrid

\begin{center}
    \textbf{\large Supplemental Material for \\
    ``Measurement-induced phase transitions in systems with diffusive dynamics''}
    
    \vspace{0.5cm}
    
    Hyunsoo Ha$^1$, Akshat Pandey$^2$, Sarang Gopalakrishnan$^3$, David A. Huse$^1$
    
    \vspace{0.2cm}
    
    $^1$\textit{Department of Physics, Princeton University, Princeton, New Jersey 08544, USA} \\
    $^2$\textit{Department of Physics, Stanford University, Stanford, California 94305, USA} \\
    $^3$\textit{Department of Electrical and Computer Engineering, Princeton University, Princeton, New Jersey 08544, USA}
\end{center}

\vspace{1cm}

\setcounter{section}{0}
\setcounter{figure}{0}
\setcounter{equation}{0}
\setcounter{page}{1}
\makeatletter
\renewcommand{\thefigure}{S\arabic{figure}}
\renewcommand{\theequation}{S\arabic{equation}}
\renewcommand{\thesection}{S\arabic{section}}
\makeatother
\setcounter{secnumdepth}{2}

\begin{center}
\begin{minipage}{0.85\textwidth}
\vspace{-0.8cm}
In this Supplemental Material, we (i) provide additional details on the classical diffusion rules and reiterate the motivation for our model, (ii) explain how the $U(1)$ symmetry precludes the volume-law phase in Clifford circuits, (iii) overview the diagnostics we used to probe the transition and the phases,  (iv) present further results near the critical point, (v) investigate the code length during the purification process, and (vi) discuss the periodic structure seen in $P^<_{S_{N/2}}(t)$ in the volume-law phase.
\vspace{1cm}
\end{minipage}
\end{center}

%%%%%%%%%%%%%%%%%%%%%%%%%%%%%%%%%%%%%%%%%%%%%%%%%%%%%%%%%%%%%%%%%%%%%%%%%%%%%%
\section{The Model: Motivation and Details}
In this section, we revisit the motivation for our model and provide additional details on the classical evolution rules for the measurer particles. We emphasize that our model investigates the nature of the measurement-induced critical point with conserved densities, extending previous models of $U(1)$ symmetric monitored circuits \cite{Agrawal_Vasseur_U1MIPT, Barratt_Potter_U1MIPTfieldtheory, chakraborty_Pixley_U1MIPT}, as we will explain below.

The $U(1)$ symmetric circuit that conserves global charge features an additional phase transition known as the ``charge-sharpening" transition, which distinguishes between the charge-sharp and charge-fuzzy phases. The MIPT occurs within the charge-sharp phase, where local charges can be identified as they effectively random walk (or classically diffuse) over large length scales, allowing the determination of the global charge by summing local contributions.

In a 1+1 dimensional qubit system, it has been reported that the critical measurement rate for the charge-conserving but otherwise Haar-random MIPT $p_c$ changes sensitively to changes in the global charge $Q$ (i.e., $d p_c/d Q \neq 0$), except when the global charge is exactly one-half and exhibits particle-hole symmetry \cite{chakraborty_Pixley_U1MIPT}. Since the critical measurement rate $p_c$ is linearly coupled to the charge, the fluctuation of charge may significantly influence the critical point, as suggested by the generalized Harris criterion discussed in the main text, which applies when $\nu < 2/d$ for $d+1$ dimensions.

As explained in the main text, we impose the conserved density through the measurement process rather than directly into the gate by introducing the ``measurers'' that evolve under the classical rules.
This allows us to investigate the MIPT critical point for large system sizes using Clifford calculations. Unlike previous models, our model does not exhibit a charge-sharpening transition. Instead, we simulate the charge-sharp phase by allowing the measurers to random-walk. A high density of measurers leads to frequent measurement events, and conversely, a low density results in fewer measurements. Our model is intended to explore how the MIPT is influenced by the space-time fluctuations of the charges. The measurement rate in our model is always linearly coupled to the density of measurers, and the generalized Harris argument remains valid for any non-trivial density of measurers (i.e., the density should neither be zero nor one). In our specific implementation, we arbitrarily set the measurer density to one-half.\\

\subsection{The classical rule for measurers: Symmetric simple exclusion process (SSEP)}
To simulate the time evolution of the conserved densities, which corresponds to the dynamics of the $U(1)$ charge in a charge-sharp phase, we adopt the well-known classical rule of the Symmetric Simple Exclusion Process (SSEP) \cite{eyink_SSEP, kipnis_SSEP, Liggett_SSEP}, where diffusive hydrodynamic equations ($z=2$) are derived from microscopic rules. 
Each measurer particle has its own clock, which rings according to a continuous Poisson process with a rate $D$. In our model, the rate is set to $D=1$, meaning the clock rings on average once during a single block of brickwork circuit. When the clock rings, the particle hops to the right or left by one site with equal probability. However, if the destination is already occupied, the movement is forbidden [illustrated in Fig.~1(a)]. For a system with $N$ qubits, we pick the number of measurers to be $N/2$, with their initial positions randomly assigned.

%%%%%%%%%%%%%%%%%%%%%%%%%%%%%%%%%%%%%%%%%%%%%%%%%%%%%%%%%%%%%%%%%%%%%%%%%%%%%%
\section{Clifford circuit with $U(1)$ symmetry}
One natural way to impose the conserved density in the model is to constrain the circuit---unitary gate and measurement---directly with $U(1)$ symmetry~\cite{Agrawal_Vasseur_U1MIPT, Barratt_Potter_U1MIPTfieldtheory, chakraborty_Pixley_U1MIPT}. However, as mentioned in the main text, the $U(1)$ symmetric Clifford circuit with measurements does not host a volume-law phase.  Instead the system purifies at long times, wandering among the computational basis product states in its statistical steady state.  We explain this in this section.

Any unitary Clifford gate changes a single Pauli string operator to a single Pauli string operator.  If a unitary Clifford gate $\mathcal{C}$ has a $U(1)$ symmetry that conserves $\sum_i \hat{Z}_i$, then $\mathcal{C} \left( \sum_i \hat{Z}_i \right)\mathcal{C}^\dagger = \sum_i \mathcal{C} \hat{Z}_i \mathcal{C}^\dagger = \sum_i \hat{Z}_i$. Therefore, each term in the sum of the left-hand side must match a term in the sum of the right-hand side, and hence, we prove the lemma below.
\begin{lemma}
    $U(1)$ symmetric Clifford circuit maps a single-site $\hat{Z}$ operator to a single-site $\hat{Z}$ operator.
    \label{lemma1}
\end{lemma}
In other words, the $U(1)$ unitary Clifford gate simply permutes the location of the single-site $\hat{Z}$ operators. Remember that the computational basis product state is stabilized by single-site $\hat{Z}$ operators on every site with signs determined by whether the spins are up or down. From the lemma above, we derive the following lemma.
\begin{lemma}
    $U(1)$ symmetric Clifford circuit maps a single computational basis product state into a single computational basis product state.
    \label{lemma2}
\end{lemma}

Now, consider an arbitrary pure stabilizer state $|\psi\rangle$ with a stabilizer group $\mathcal{G}$, such that for any $g\in\mathcal{G}$, $g|\psi\rangle = |\psi\rangle$. If one calculates the expectation value of the single-site $\hat{Z}$ operator at site $r$, $\langle \psi | \hat{Z}_r | \psi \rangle$, it should always be $\pm 1$ or zero as below:
\begin{align}
    \langle \psi | \hat{Z}_r | \psi \rangle 
    = \Tr\left(\hat{Z}_r|\psi\rangle\langle\psi|\right)
    = \Tr\left(\hat{Z}_r \frac{1}{|\mathcal{G}|}\sum_{g\in\mathcal{G}} g  \right)
    =\frac{1}{|\mathcal{G}|}\sum_{g\in\mathcal{G}}\Tr\left(\hat{Z}_r g \right)
    =\begin{cases}
		1, & \hat{Z}_r \in \mathcal{G}\\
        -1, & -\hat{Z}_r \in \mathcal{G}\\
        0, & \textit{otherwise}
	\end{cases}
\end{align}
as the trace of the nontrivial Pauli string is always zero while the trace of the identity matrix is $2^N$, and we used $|\mathcal{G}|=2^N$. 
If one writes the stabilizer state in terms of the computational basis states as $|\psi\rangle = \sum_{\{s_i\}}W_{\{ s_i\}}|\{ s_i\}\rangle$, where each $s_i=\pm 1$, on sites ($r$) where $\langle \psi|\hat{Z}_r|\psi\rangle=\pm 1$ only basis states with that $s_r$ are present. Hence, the total number of basis states used to represent $|\psi\rangle$ is upper-bounded by $2^n$, where $n$ is the number of sites with $\langle\psi|\hat{Z}_i|\psi\rangle=0$.

Every time when a $U(1)$-symmetric Clifford unitary gate acts on the stabilizer state, the number of computational basis states in its expansion remains the same (Lemma~\ref{lemma2}).  If $\langle \hat{Z}_r \rangle=\pm 1$ and then we measure $\hat{Z}_r$, the number of computational basis states remains the same.  However, if we measure a site where $\langle \hat{Z}_r \rangle=0$, the wavefunction then ``collapses'' to either spin up or spin down, and the upper bound on the number of basis states decreases by a factor of two. Therefore, the initial basis state number upper-bounded by $2^n$ is reduced to one after all sites where $\langle \hat{Z}_r \rangle=0$ are measured, which will typically take a time of order $\sim (\log{N})/p$ or less, where $p$ is the measurement rate.  Essentially the same thing happens if we instead start with a mixed stabilizer state. Thus we prove the proposition below:
\begin{proposition}
    $U(1)$ symmetric Clifford circuit with finite rate of $\hat{Z}$ measurement always purifies at long time to a computational basis product state.
\end{proposition}

\section{Diagnostics Overview}
\subsection{Entanglement probes}
We use entanglement probes to investigate the phase diagram of our model. Starting from a pure state (mostly using product initial states, but we can also start from random pure states), we characterize distributions of several quantities over the course of the dynamics, as well as in the long-time steady state.  Specifically, we consider the (i) half-chain entanglement entropy $S_{N/2}$, (ii) antipodal mutual information: AMI, and (iii) tripartite mutual information $\mathcal{I}_3$, defined as follows:
\begin{align}
    S_{N/2} &\equiv S_A;\\
    \mathrm{AMI} &\equiv S_A + S_B - S_{A\cup B};\\
    \mathcal{I}_3 &\equiv S_A + S_B + S_C - S_{A \cup B} - S_{B \cup C} - S_{A \cup C} + S_{A\cup B\cup C}.
\end{align}

\begin{figure}[h!]
    \centering
    \includegraphics[width=0.6\textwidth]{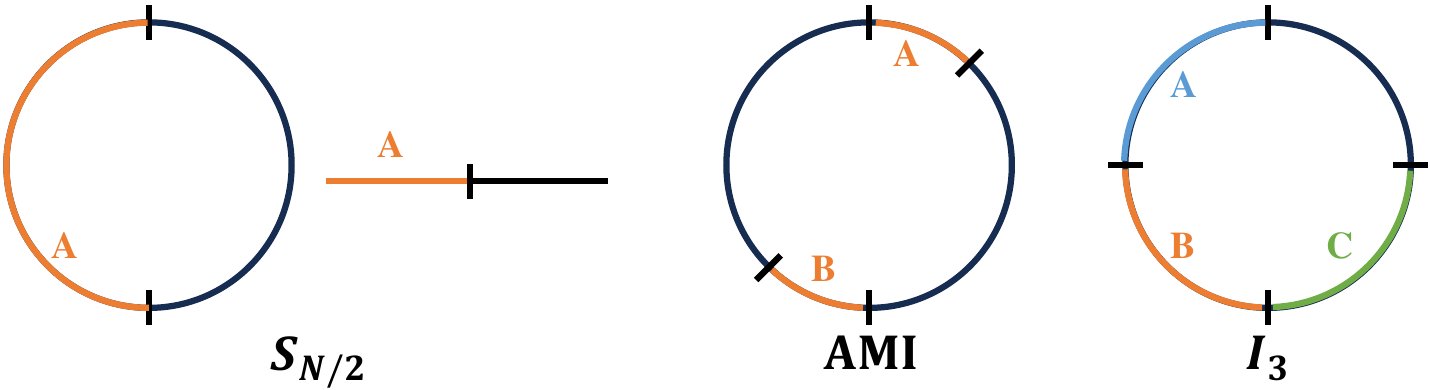}
    \caption{Illustrations of partitions into subsystems involved in various entanglement probes. (Left) Half-chain bipartite entanglement $S_{N/2}$ from the periodic and open boundary condition. (Middle) Antipodal mutual information (AMI) between subregions $A$ and $B$ with length $N/8$ and centers $N/2$ apart. (Right) Tripartite mutual information $\mathcal{I}_3$ where $|A|=|B|=|C|=N/4$.}
  \label{fig_diagnostics_entanglement}
\end{figure}

The bipartite entanglement $S_{N/2}$ is the simplest probe diagnosing the volume-law and the area-law phase, characterized by $S_{N/2}\sim N$ and $S_{N/2} \sim O(1)$, respectively, in the steady state. Also, with a product initial state in the volume-law phase, $S_{N/2}(t)$ increases linearly in $t$ before it saturates to an order-$N$ value. 
We additionally investigate $S_{N/2}$ with open boundary conditions while exploring the Griffith-like effects because now, a single cut defines a bipartition, and it better captures the rare-region effect.
However, $S_{N/2}$ is sensitive to entanglement on all length scales. 
AMI and $\mathcal{I}_3$ are, in some situations, nicer objects to work with, since they are sensitive only to long-range entanglement.

In particular, we use the antipodal mutual information AMI to estimate the location of the critical point \cite{Li_Chen_Fisher}. We expect $\mathcal{I}_{AB}\rightarrow0$ in both volume-law and area-law phases in the thermodynamic limit, so for finite systems AMI has a peak near the critical point, which can be found without making any scaling assumption. 

We also investigated the tripartite mutual information $\mathcal{I}_3$, a probe that has been found to be useful for precisely characterizing the spacetime-uncorrelated conventional MIPT \cite{Zabalo_Pixley_diagnostics}. It is asymptotically zero in the area-law phase and is negative and proportional to $N$ at the volume-law phase. However, the value of $\mathcal{I}_3$ is sensitive to the specific circuit realization, and the probability distribution $\mathcal{P}[\mathcal{I}_3]$ exhibits a fat tail under strong space-time correlation, i.e., quenched randomness \cite{Zabalo_Pixley_infiniterandomnessMIPT, shkolnik_gazit_quasiperiodicMIPT}. In such scenarios, the average value of $\mathcal{I}_3$ is not representative of the distribution, and it may be more informative to study the quantity $\mathcal{P}[\mathcal{I}_3=0]$, which we also explore.

\subsection{Purification dynamics}
Together with the entanglement probes, we investigate purification dynamics \cite{Gullans_Huse_purification}, establishing a direct connection between space and time. 
Specifically, we prepare a fully mixed initial state $\hat{\rho}_i = \mathbb{\hat{I}}/2^N$ and track the von Neumann entropy $S(t) = -\tr(\hat{\rho}_t\log_2\hat{\rho}_t)$ over time.
The purification time $\tau_P$ is when the system is completely purified, so that $S(t) = 0$ for all $t\geq \tau_P$.
The volume-law to area-law phase transition in the conventional MIPT is also a transition from a ``mixed'' phase with exponentially long purification times $\log{\tau_P} \sim N$, to a ``purifying'' phase where $\tau_P$ grows only as $\log{N}$.
At the critical point, $\tau_P\sim N^z$ with $z=1$ for spacetime-uncorrelated measurements, while $\log \tau_P \sim N^\psi$ (so that $z = \infty$) for quenched spatial randomness.

\subsection{Contiguous code length}
\label{contiguous_cod_length}
The contiguous code length \cite{Bravyi_2009, Gullans_Huse_purification, Ippoliti_Khemani_measurementonly} shows where the logical operator associated with the state is located. 
In this work, we calculate the contiguous code length when there is a single logical pair (or, equivalently, $(N-1)$ stabilizers).
A single logical pair can be prepared in the purification protocol starting from the maximally mixed state, waiting until exactly one bit remains unpurified. 
Calling the logical operators $\hat{\ell}_X$, $\hat{\ell}_Y$, and $\hat{\ell}_Z$, where the logicals follow the same algebra as Pauli $\hat{X}$, $\hat{Y}$, and $\hat{Z}$ operators respectively, we can 
find the shortest contiguous interval that supports each logical operator, with the respective lengths $d_\alpha$ defined as
\begin{align}
    d_\alpha = \min_A\{ |A|: \exists~\hat{\ell}_\alpha \mathrm{~supported~in~}A \},\qquad \alpha\in\{X,Y,Z\}.
\end{align}
where $A$ denotes an interval within the chain.

Initially, we entangle the logical pair with a reference ancilla (labeled $a$) to form a Bell pair so that $\hat{X}_a\otimes\hat{\ell}_X$ and $\hat{Z}_a\otimes\hat{\ell}_Z$ are the stabilizer generators of the ancilla-extended system. 
We then run the circuit solely on the system, so the time-evolved logicals are $\hat{X}_a\otimes \hat{\ell}_X(t)$, $\hat{Y}_a \otimes \hat{\ell}_Y(t)$, and $\hat{Z}_a \otimes \hat{\ell}_Z(t)$.
The system gets purified at time $\tau_P$ once the measurements coincide with the logical operator (assuming $\hat{\ell}_X(\tau_P)$ is measured), leading to the disentanglement of the ancilla from the system. The stabilizers at the purification time are $\mathbb{\hat{I}}_a\otimes\hat{\ell}_X(\tau_P)$ and $\hat{X}_a\otimes\mathbb{\hat{I}}$.

To independently capture the code length for each logical component, we adopt the so-called clipped gauge \cite{Nahum_PRXentanglementgrowth,Li_Chen_Fisher}. Each stabilizer generator has 
a left and a right ``end'' of nontrivial content, so a pure state of length $L$ has a total of $2L$ ends. The clipped gauge imposes the condition that exactly two ends, originating from either different generators or the same one, are positioned at every site. Such a representation allows efficient calculation $S_A$ of the contiguous interval $A$ by counting the string of generators with one end inside the interval and the other outside.

When applying the clipped gauge to a system with a single logical pair coupled to an ancilla as a Bell pair (so the total length is $N+1$), we find two stabilizer generators with nontrivial content starting from the first site (the ancilla's location) and ending at $r_1+1$ and $r_2+1$, respectively ($r_1 \leq r_2$). The mutual information between the ancilla and the system within an interval from $1$ to $r$ (denoted as $A_{[1,\cdots,r]}$) is determined as follows:
\begin{align}
    \mathcal{I}(\mathrm{ancilla};~A_{[1,\cdots,r]}) = 
    \begin{cases}
        0 \quad (r<r_1), \\
        1 \quad (r_1\leq r < r_2), \\
        2 \quad (r_2 \leq r).
    \end{cases}
\end{align}
If the shorter stabilizer generator has Pauli operator $\hat{\alpha}_a$ at the ancilla, then $A_{[1,\cdots,r]}$ with $r_1\leq r < r_2$ supports the logical operator $\hat{\ell}_\alpha$, while the other two are absent. For intervals longer than $r_2$, they include all three ($\hat{\ell}_X(t)$, $\hat{\ell}_Y(t)$, and $\hat{\ell}_Z(t)$) logical operators:
\begin{align}
    A_{[1,\cdots,r]}\textrm{~includes~} 
    \begin{cases}
        \textrm{no~logicals} \quad &(r<r_1), \\
        \hat{\ell}_\alpha \quad &(r_1\leq r < r_2), \\
        \hat{\ell}_X, \hat{\ell}_Y, \textrm{and~} \hat{\ell}_Z &(r_2 \leq r).
    \end{cases}
\end{align}

To determine the code length of logicals in a periodic chain, we repeatedly compute in the clipped gauge while keeping the ancilla position fixed, cyclically permuting the system. 
During each iteration, we obtain $r_1$ and $r_2$, and identify which logical component is supported on length $r_1$ (flagged with $\hat{\alpha}_a$). Subsequently, we designate length $r_1$ to $\alpha$ and $r_2$ to 
$\{X,Y,Z\}\setminus\{\alpha\}$, 
comparing these values across iterations to identify the shortest length and its corresponding interval for all $X$, $Y$, and $Z$ logical components.

We note that the shortest interval supporting each logical may not be unique, but the shortest length is well defined. Furthermore, since logicals with different components anticommute, the intervals supporting different logicals should always overlap on at least one site.

%%%%%%%%%%%%%%%%%%%%%%%%%%%%%%%%%%%%%%%%%%%%%%%%%%%%%%%%%%%%%%%%%%%%%%%%%%%%%%
\section{Results near the critical point}
\subsection{Determining the critical point}
 \begin{figure}[h]
    \centering
    \includegraphics[width=\textwidth]{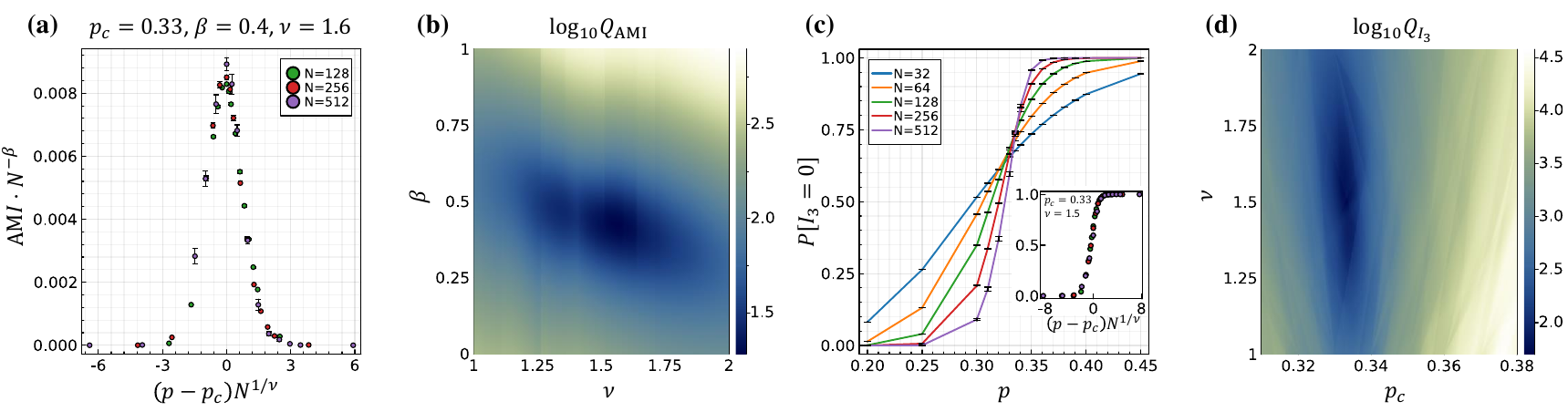}
    \caption{
    (a) Re-scaled AMI for large system sizes ($N\geq128$)
    (b) $\log_{10} Q_{\mathrm{AMI}}$, an indicator of the quality of the one-parameter scaling ansatz Eqn.~\ref{eqn:AMI_scaling} for different values of $\beta$ and $\nu$. Smaller values represent the better collapse, with $\log_{10} Q_{\mathrm{AMI}} = 0$ indicating the best result.
    (c) $\mathcal{P}[\mathcal{I}_3=0]$ at the steady value for various measurement rates and system sizes. Inset shows the finite-size scaling of $\mathcal{P}[\mathcal{I}_3=0]$ for large system sizes ($N\geq128$).
    (d)  $\log_{10} Q_{I_3}$ shows the quality of the scaling ansatz Eqn.~\ref{eqn:PI3_scaling} for different values of $\nu$ and $p_c$.
     }
  \label{fig_critical point}
\end{figure}

As explained in the main text, we rely on the steady-state antipodal mutual information to determine the critical measurement rate, constrained to be in between the interval $p_c \in (0.31,0.35)$. Solely focusing on the peak of AMI suggests $p_c\approx0.33$.
The increasing peak for larger system sizes is a notable difference from the conventional space-time uniform MIPT, where AMI has a constant value at the critical point. The evident increasing peak indicates the absence of conformal invariance in our model. 

For the steady-state values, circuits are applied until time $t=5t_{\mathrm{classical}}$, where $t_{\mathrm{classical}}=(N/2\pi)^2$ represents the relaxation time on the largest length scale of the classical SSEP particles.
We confirmed that these values saturate within our chosen circuit depth.
The results are averaged over $5\times10^3$ to $3\times10^4$ circuit realizations. For each realization, we also averaged over the last 1000 time units to achieve better resolution, as the entanglement values for stabilizer states are discrete integers. The number of samples was adjusted based on the uncertainties in order to reduce error bars and ensure that the data points could be clearly distinguished from one another.
As an initial state, we start with a random product stabilizer state (with no initial entanglement). Additionally, we confirmed that the entanglement probes saturate to the same values even when we begin with a pure random stabilizer state (a maximally entangled state). This demonstrates that the final results are independent of the initial state configuration.

Adopting the peak to be $p_c$, next, we attempt to extract the correlation-length exponent $\nu$ from the following scaling ansatz
\begin{align}
\mathrm{AMI} \sim N^\beta f(N^{1/\nu}(p-p_c))
\label{eqn:AMI_scaling}
\end{align}
where $f$ is the scaling function. As shown in Fig.~\ref{fig_critical point}(a), we use $N$=128, 256, and 512 to avoid being distracted by the smaller system sizes, which are unlikely to scale properly, and find $\nu = 1.6$. However, the data collapse is poor, as we quantified in Fig.~\ref{fig_critical point}(b) with the objective function $Q_{\mathrm{AMI}}$ that determines the quality of the scaling ansatz for different values of $\beta$ and $\nu$ \cite{scaling,Zabalo_Pixley_diagnostics}.

We additionally present results on the tripartite mutual entanglement $\mathcal{P}[\mathcal{I}_3=0]$ in Fig.~\ref{fig_critical point}(c). The pronounced drift towards larger $p_c$ for larger system sizes becomes evident as we focus on the intersection of the two graphs, indicating significant finite-size effects. We again attempt to extract the correlation length exponent from the following ansatz
\begin{align}
P[I_3 = 0] \sim g(N^{1/\nu}(p-p_c))
\label{eqn:PI3_scaling}
\end{align}
with a scaling function $g$, and get $\nu=1.5$ as shown in the inset of Fig.~\ref{fig_critical point}(c). Fig.~\ref{fig_critical point}(d) shows the objective function for the scaling ansatz Eqn.~\ref{eqn:PI3_scaling}.
From both AMI and $P[I_3=0]$ observables, we extract $\nu=1.6(5)$; however, the large uncertainties presented in Fig.~\ref{fig_critical point}(b,d) and the strong finite-size effects shown in Fig.~\ref{fig_critical point}(c) that everything may still strongly drift hinder the accurate characterization of the scaling relations.

%%%%%%%%%%%%%%%%%%%%%%%%%
\subsection{Entanglement growth at the critical point}
\begin{figure} [h]
    \centering
    \includegraphics[width=\textwidth]{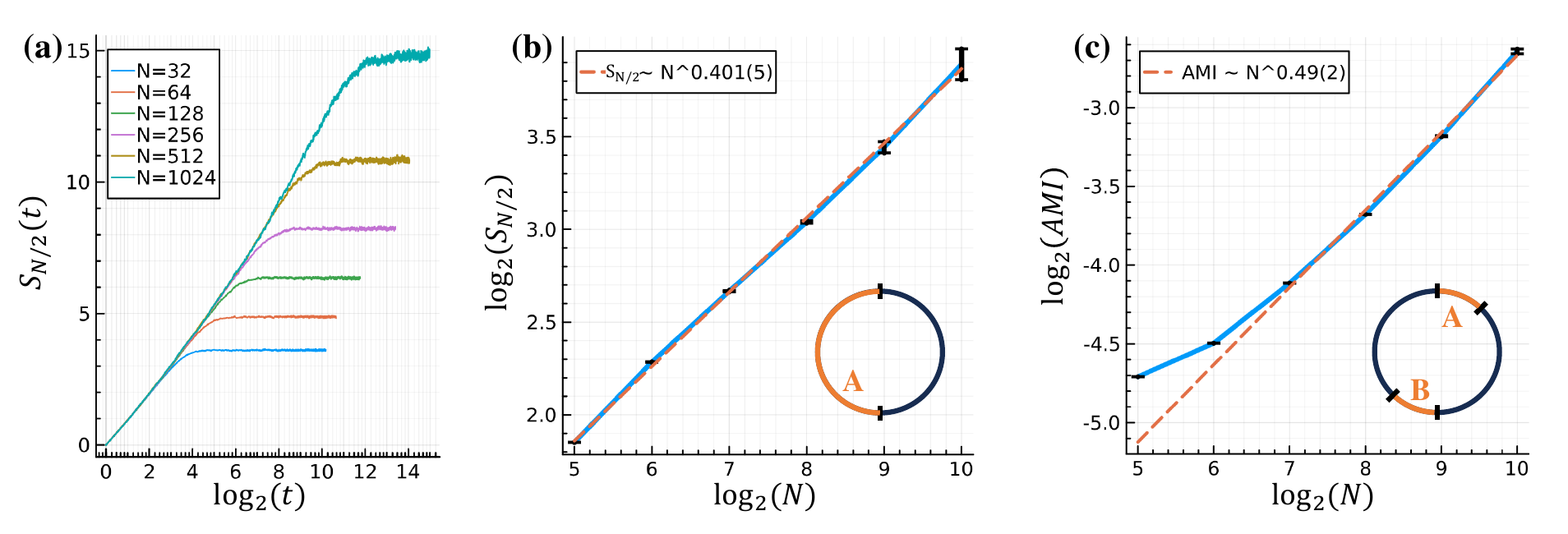}
    \caption{Entanglement probes at $p=0.33$. (a) $S_{N/2}(t)$ shows a logarithmic trend. (b,c) Steady-state properties are presented with (b) $S_{N/2}$ and (c) AMI, averaged over samples, showing an algebraic increase over sample sizes.}
  \label{fig_entanglement_result}
\end{figure}
Once the critical point is identified, we investigate the entanglement growth. While investigating such dynamics, we present for a shorter time scale (scaling linearly with $N$) after confirming that the $S_{N/2}$ saturates within this circuit depth. 
Fig.~\ref{fig_entanglement_result} explicitly present the result for $p=0.33$, indicating a logarithmic trend [Fig.~\ref{fig_entanglement_result}(a)].
A notable distinction is observed in the steady-state entanglement, as depicted in Fig.~\ref{fig_entanglement_result}(b,c), where the relation between $S_{N/2}$ and AMI over various system sizes demonstrates an algebraic relation $S_{N/2}\sim N^{\psi'}$ with $\psi'\approx0.4$. This behavior resembles the quenched randomness critical point, which has $S_{N/2}\sim N^{1/2}$, and the power-law scaling was also observed in the system with quantum and classical circuits coupled together\cite{klocke_moore_buchhold2024critical}.

%%%%%%%%%%%%%%%%%%%%%%%%%
\subsection{Purification dynamics at the critical point}
\begin{figure} [h]
    \centering
    \includegraphics[width=0.7\textwidth]{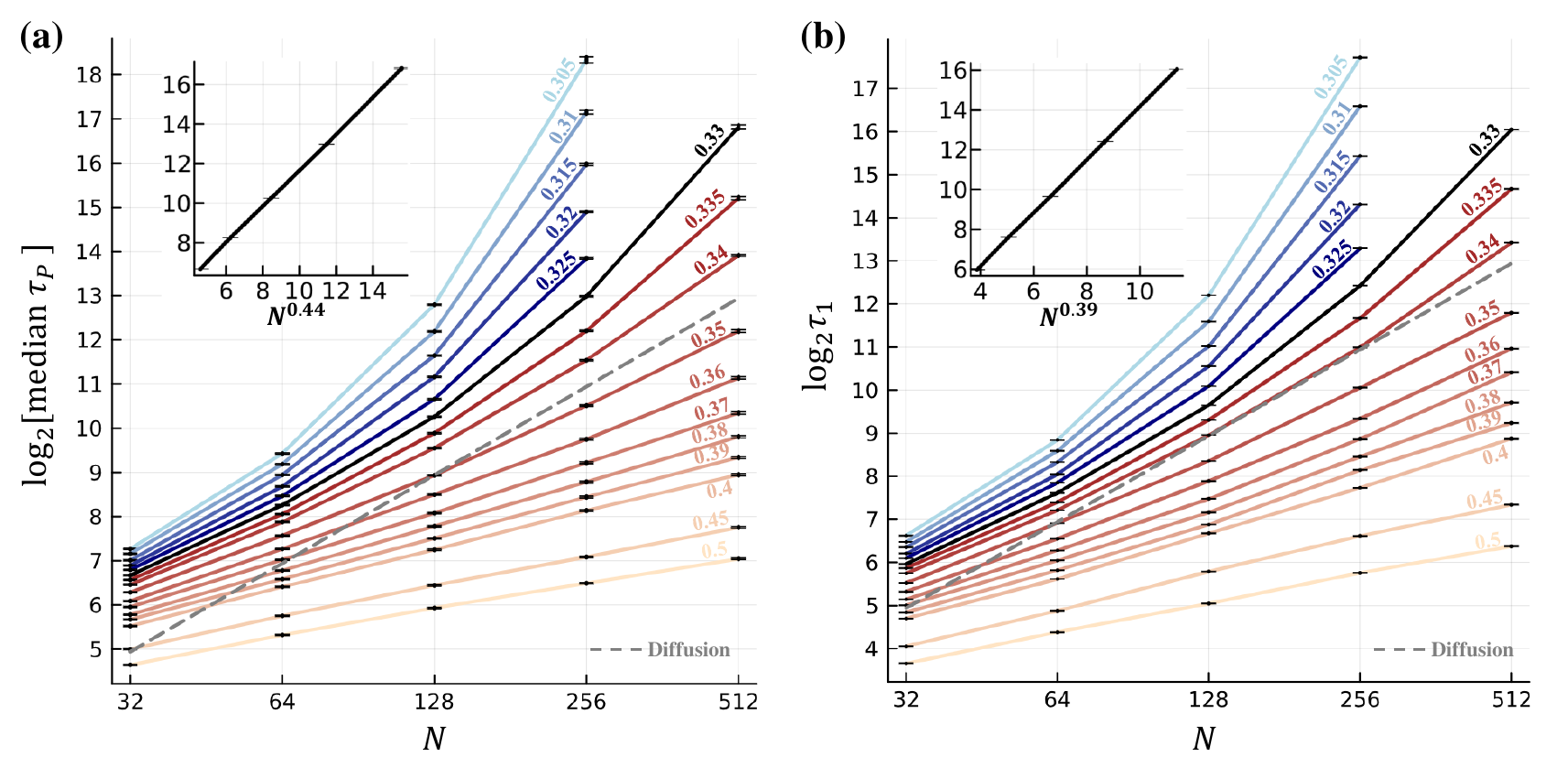}
    \caption{(a) Purification time $\tau_P$. The median over samples is presented as a representative of the typical samples. The inset shows activated dynamical scaling at $p=0.33$, with the x-axis rescaled to $N^{0.44}$. (b) The lifetime of the last logical bit $\tau_1$. The inset shows $p=0.33$ with the x-axis rescaled to $N^{0.39}$. The measurement rates ($p$) are indicated in both graphs with numbers on each plot with different colors. The gray dotted line is the relaxation time scale for the diffusive measurers with $z=2$.}
  \label{fig_purification_result}
\end{figure}

As explained in the main text, we explore the purification dynamics to find a direct connection between space and time at the critical point. As shown in Fig.~\ref{fig_purification_result} (a), it is clear that the new critical point has $z>1$, indicating a different universality class from the conventional MIPT. From the constrain of $p_c\in(0.31,0.35)$ based on the AMI results, we can bound the dynamical exponent $z\geq2$, and there is a signature of diverging dynamical exponent $z\rightarrow\infty$ in the thermodynamic limit if $p_c=0.33$.

The long-lived behavior of the last logical pair also captures the extraordinarily long timescale.
Starting from the fully mixed state with entropy $S(0)=N$, the entropy monotonically decreases over time. Furthermore, the entropy is always an integer, due to the Clifford character of our circuits. Therefore, the moment when the entropy remains precisely one bit --- the ``last logical'' --- is well defined.
The probability distribution of the last logical's survival time turns out to have an exponential tail near the critical point.
One can extract the last logical's lifetime $\tau_1$ by fitting the exponential tail. This lifetime has roughly the same behavior as the purification times, as shown in  Fig.~\ref{fig_purification_result}(b).

\begin{figure}
    \centering
    \includegraphics[width=0.8\textwidth]{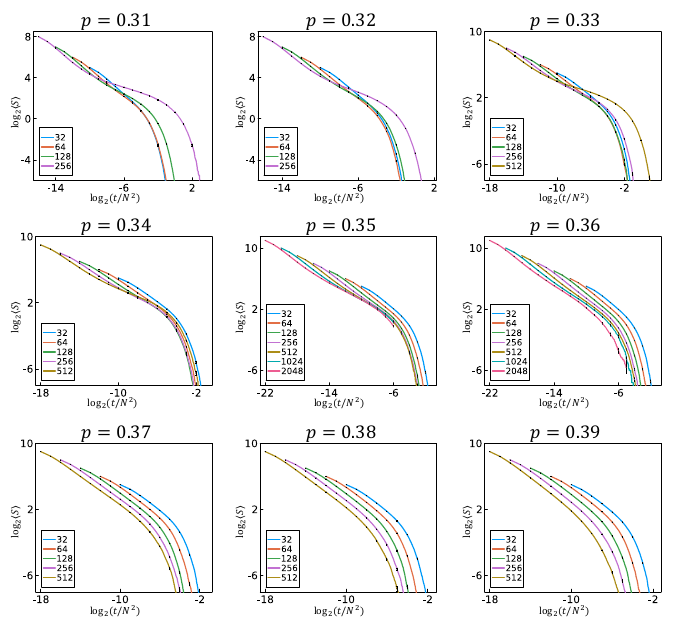}
    \caption{Entropy from the purification dynamics, presented over re-scaled time $t/N^2$.}
  \label{fig_entropy_scaling}
\end{figure}

We also present how the entropy $S(t)$ decays over time (scaled as $t/N^2$) at different measurement rates [Fig.~\ref{fig_entropy_scaling}]. 
We investigate the possibility of the dynamical exponent being the same as that of the measurers' SSEP viz.  $z=2$. For $p=0.33$, the entropy develops a plateau for larger system sizes, clearly showing $z>2$, while for the slightly higher measurement rate ($p=0.35$), the plots seem to collapse for various system sizes. Collapsing plots near $p\approx0.35$ with re-scaled time $t/N^2$ suggests a scenario, discussed in the main text, where $p_c$ is larger than the $p \approx 0.33 $ which has heretofore been the focus of our discussion, with $z=2$ dynamical scaling at criticality. Even at these rather large system sizes, we believe our model suffers from significant finite-size corrections to scaling. We leave further investigation to characterize the critical behavior for future work.

%%%%%%%%%%%%%%%%%%%%%%%%%
\section{Purification of the last logical bit}

In this section, we focus on the $z\rightarrow\infty$ scenario and provide a qualitative argument focusing on the code length of the last logical bit.
Naively, if one considers a rare region of length $\ell$ with a relatively low density of measurers. The reduced density in this rare region leads to a lower measurement rate, allowing logical operators to remain secure inside. However, due to the diffusive nature of measurers, the rare region has a finite lifetime of $O(\ell^2)$, and the logicals will be exposed to a higher measurement rate afterward, suggesting that the dynamical exponent should be bounded by $z\leq2$. 

Contrary to expectations, however, the logical can hop between rare regions, as illustrated in Section~\ref{Appendix: visualizing logical}. This occurs when one logical operator (denoted as $\hat{X}_L$) increases in size while the other logical operator ($\hat{Z}_L$) maintains its length. The shorter logical appears to tunnel between rare regions, taking advantage of the sufficiently long pair. This mechanism may be necessary so that the logical can persist even after the rare region dissipates, allowing the possibility of $z>2$.

\subsection{Contiguous code length of the last logical}

\begin{figure}[!ht]
    \centering
    \includegraphics[width=\textwidth]{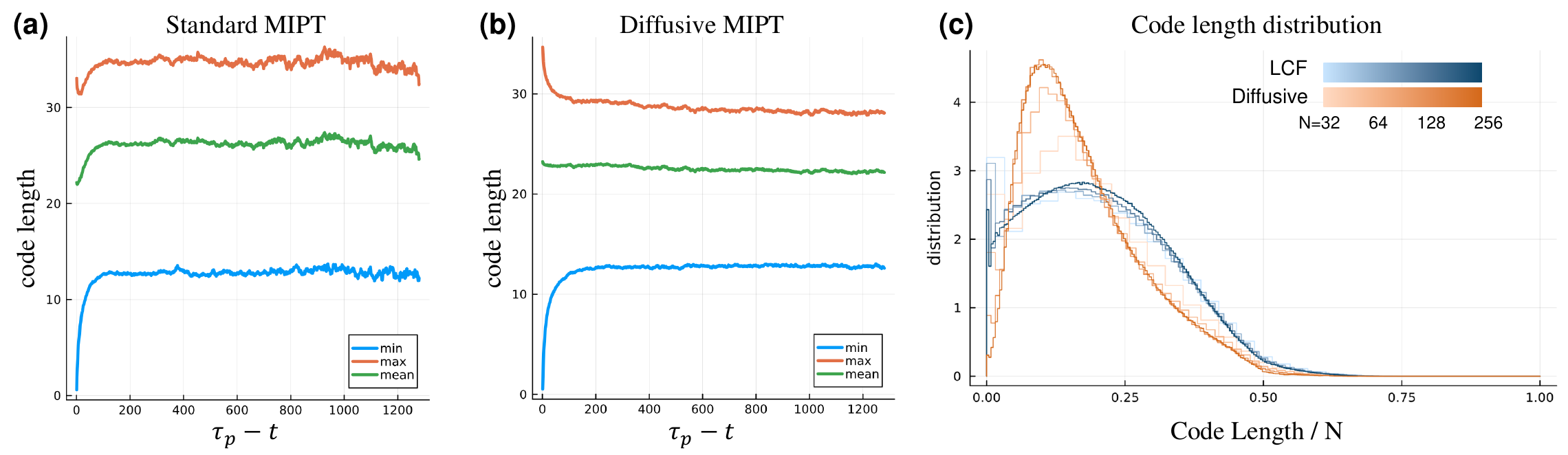}
    \caption{(a,b) Contiguous code length of the logical before it gets purified. (a) Standard MIPT in Clifford circuits and (b) our diffusive measurer model at $p=0.33$ are presented. The x-axis displays $\tau_P - t$, the remaining time before being purified. All calculations are performed in the system length $N=128$. (c) Probability distribution of the contiguous code length. Both standard MIPT \cite{Li_Chen_Fisher} at the critical point and diffusive model at $p=0.33$ are presented. The system size increases as the plot gets darker. The code length scales proportional to the system size.}
  \label{fig_codelength}
\end{figure}

Fig.~\ref{fig_codelength}(a,b) illustrates the contiguous code length of the last logical pair for the standard MIPT in Clifford circuits \cite{Li_Chen_Fisher} at the critical point, and in our model with $p=0.33$. The x-axis denotes $\tau_P - t$, the time until purification.
The code length of three nontrivial logical operators $\hat{X}_L$, $\hat{Z}_L$, and $\hat{Y}_L=i\hat{Z}_L\hat{X}_L$ was calculated, and we averaged the shortest, longest, and mean values over samples.

The shortest logical operator consistently shrinks to zero (single-site occupation) preceding purification in both models.
However, the distinction is clear when one focuses on the longest logical operator. In contrast to the standard model, the diffusive MIPT model exhibits an increase in the length of the longest logical operator. This observation supports our previously addressed mechanism.

The long-lived behavior is also reflected in the contiguous code length probability distribution, shown in Fig.~\ref{fig_codelength}(c). In the conventional MIPT, the probability is finite at the zero-length, while the diffusive model produces a clear peak near $0.1N$, with the distribution vanishing at zero code length. This implies the code length remains extensive most of the time and has a much smaller chance of going to zero.

\subsection{Hyperuniformity of measurers when the last logical purifies}
\begin{figure}[!ht]
    \centering
    \includegraphics[width=0.9\textwidth]{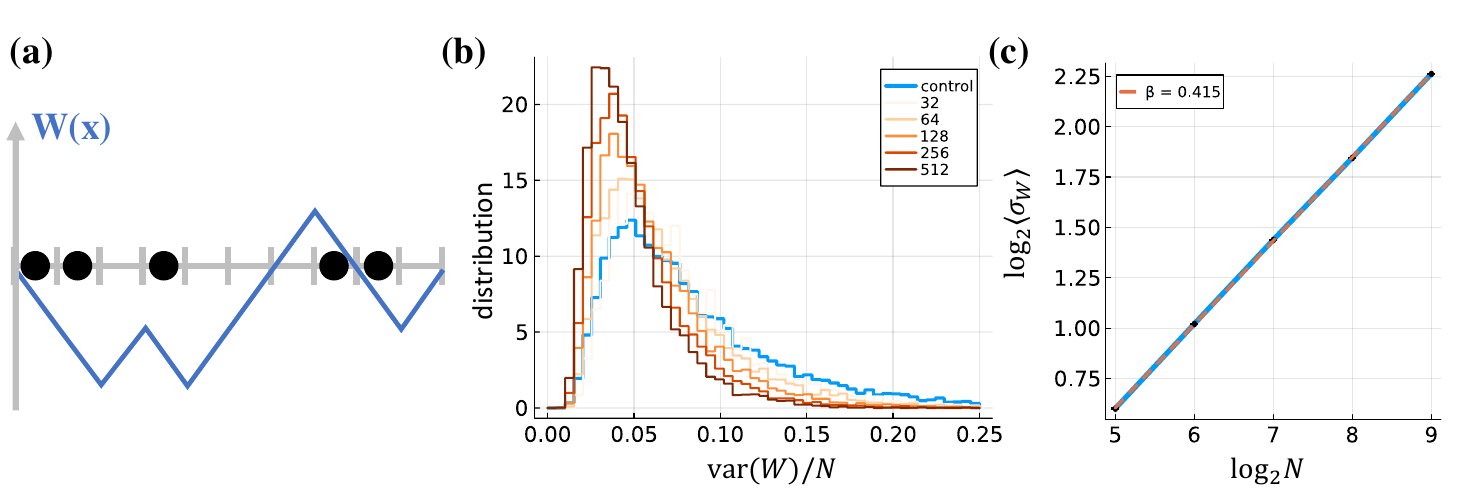}
    \caption{ (a) Schematic of the height function $W(x)$. (b) Probability distribution of the $W(x)$ variance sampled when the ancilla is disentangled from the system. Compared to the control group (blue) from randomly distributed measurers, the probability distribution of $\mathrm{var}(W)$ gets sharper, and the peak shifts towards zero. (c) Hyperuniformity of the measurers with wandering exponent $\beta=0.415$. The sample-averaged standard deviation of $W$ and the system sizes have clear algebraic relations, as shown in the log-log plot.}
  \label{fig_W(x)}
\end{figure}
From the observations on code length above, we qualitatively and quantitatively analyzed that 
(1) the contiguous code distance is proportional to the system size for the typical times, and 
(2) even very short logical operators with $O(1)$ length may be secured by the rare region with a lower density of measurers, and can grow longer. 
Based on these observations, we propose that the purification lifetime is controlled by events when the spatial rare regions are suppressed system-wide. This leads to ultra-slow dynamics with purification time characterized by stretched exponential of system size.

As a measure of the uniformity of classical particles, we define a height function $W(x)$:
\begin{align}
    W(x) = \sum_{r=1}^{x} (-1)^{n(r)},
\end{align}
with $W(0)=0$, and $n(r)$ being the occupation number of measurers at $r$. We have $W(N)=W(0)$, since the number of measurers is exactly half the system size. 
If measurers are randomly distributed, which is true for the typical distribution of SSEP particles, the standard deviation of the height function scales $\sigma(W)\sim N^{1/2}$.
However, the measurer locations are not random at the moment when the system gets totally purified, and hyperuniformity \cite{Torquato_hyperuniformreview} characterized by the wandering exponent $\beta$, is observed at this particular time: $\sigma(W) \sim N^\beta$, with $\beta=0.415$ [Fig.~\ref{fig_W(x)}(c)]. Fig.~\ref{fig_W(x)}(b) also shows the distribution $\mathrm{var}(W)$ at the purification time is clearly different from the distribution for randomly distributed measurers (blue line).
The probabilistic chance of measurers to locate hyperuniformly is approximately $\mathrm{exp}(N^{1-2\beta})$ (ignoring constants). This heuristic mechanism supports a picture of ultra-slow dynamics with $\psi\approx 1-2\beta$.

\section{In-periodic structure of $P^<_{S_{N/2}}(t)$}
\begin{figure}[!ht]
    \centering
    \includegraphics[width=\textwidth]{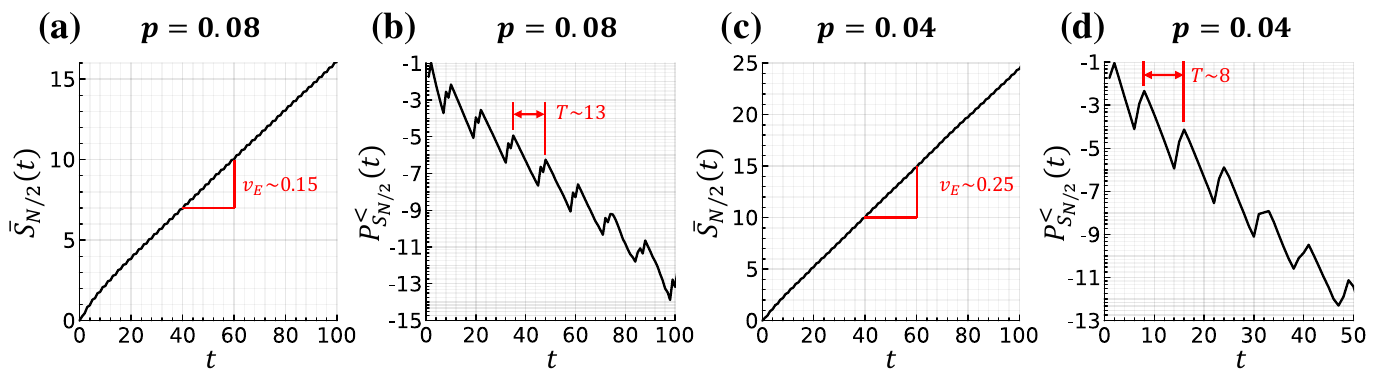}
    \caption{(a,b) Results from an open-boundary chain of length $N=1024$ and measurement rate $p=0.08$. The measurements are spacetime-independent and identically distributed, and the measurement rates are low enough to be well inside the volume-law phase. 
     In (a), we observe ballistic growth of bipartite entanglement from a single cut at the middle, indicating a velocity of entanglement propagation of approximately $v_E\sim0.15$. In (b), the nearly periodic structure in $P^<_{S_{N/2}}(t)$ has a period of approximately $T\sim13$, which aligns with the expected period of $2/v_E$, as explained in the text.
     (c,d) Here, we present another demonstration with different measurement rates, $p=0.04$, and find $v_E\sim0.25$ and $T\sim8$, consistent with the anticipated relationship. }
  \label{fig_supp_periodic}
\end{figure}

The tail distribution of entanglement growth effectively captures Griffiths-like effects within the volume-law phase, arising from rare regions of low measurer density with finite lifetimes due to particle diffusion. We define $P^<_{S_{N/2}}(t)$ as the probability of bipartite entanglement being suppressed by more than half the average over samples for a single cut. In our analysis, we observed distinct scaling behaviors for $P^<_{S_{N/2}}(t)$ across three classes of models with different spacetime correlations of measurements.

However, alongside the anticipated scaling relations, we observed the emergence of a nearly periodic structure over time (see Fig.~3 in the main text). This periodicity is a characteristic feature of the Clifford circuit, where all $S_{N/2}(x,t)$ values are integers for each run. Unlike individual samples of $S_{N/2}$, the mean value $\bar{S}_{N/2}$, and consequently $\bar{S}_{N/2}/2$, are not necessarily integers. As $\bar{S}_{N/2}$ increases by 2 units, the range of integers considered for $P^<_{S_{N/2}}(t)$ expands as a step function over time.

This characteristic is readily apparent in the spacetime-uncorrelated MIPT \cite{Li_Chen_Fisher}, where entanglement grows ballistically, i.e., $\bar{S}_{N/2}(t) = v_E t$. Consequently, the range of counting for $P^<_{S_{N/2}}(t)$ increases every $2/v_E$ periods, leading to the observed periodic structure over time. Figure~\ref{fig_supp_periodic} illustrates this feature for measurement rates $p=0.08$ and $p=0.04$, where the period is indeed determined by $2/v_E$.

\section{Appendix: visualizing the last remaining logical bit}
\label{Appendix: visualizing logical}
\begin{figure}[!ht]
    \centering
    \includegraphics[width=\textwidth]{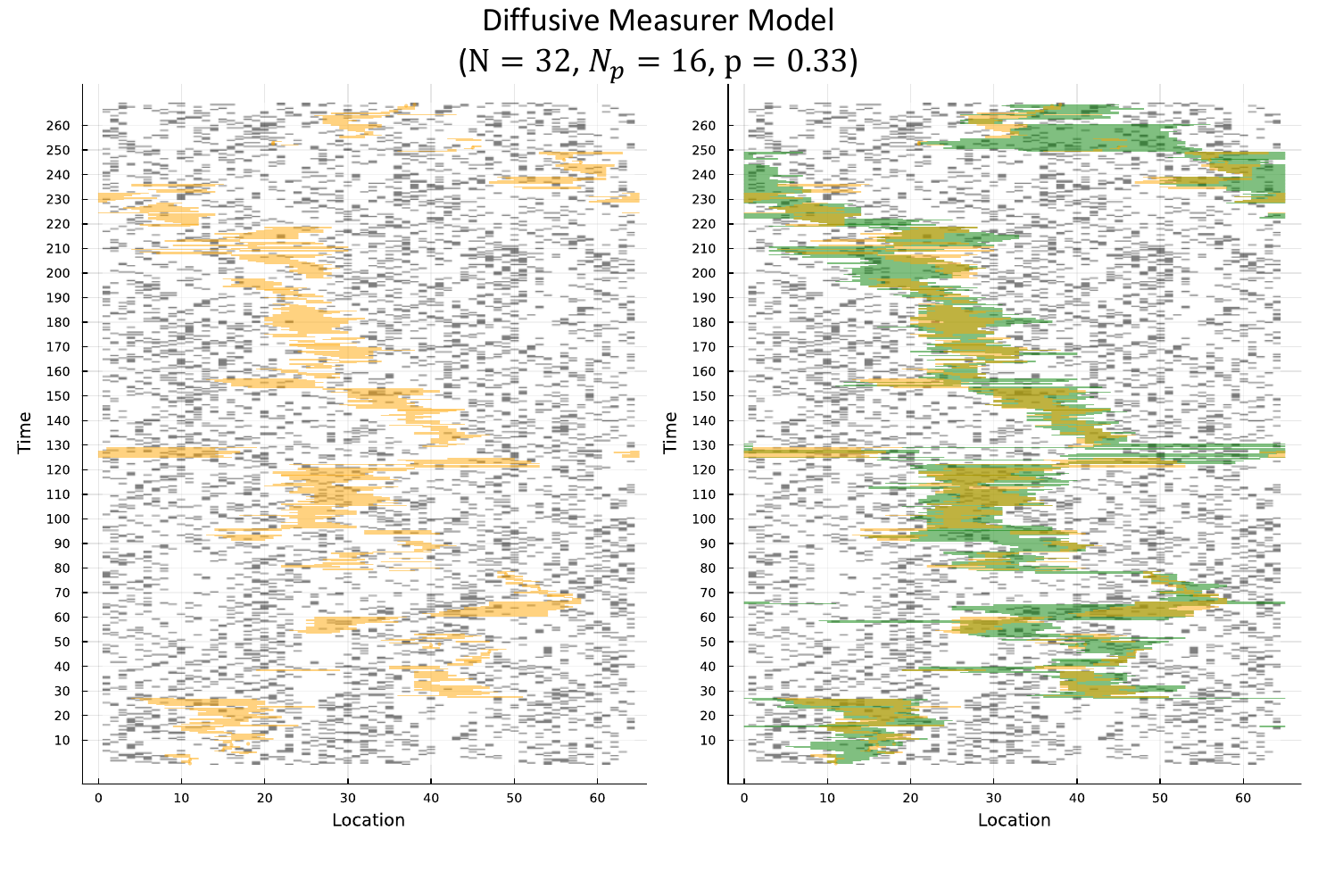}
    \caption{The shortest logical operator (left) and both the longest and the shortest logical pair (right) in our diffusive model.}
  \label{fig_logical_diffusive}
\end{figure}

We here visualize the last remaining logical bit for $p=0.33$ with the contiguous code length and the corresponding interval described in Section \ref{contiguous_cod_length}. In Fig.~\ref{fig_logical_diffusive}, we show the shortest(longest) logical operator among $\hat{\ell}_X$, $\hat{\ell}_Y$, and $\hat{\ell}_Z$ at individual time with yellow(green). The background gray shows the space-time location of the measurements.

\begin{figure}[!ht]
    \centering
    \includegraphics[width=0.7\textwidth]{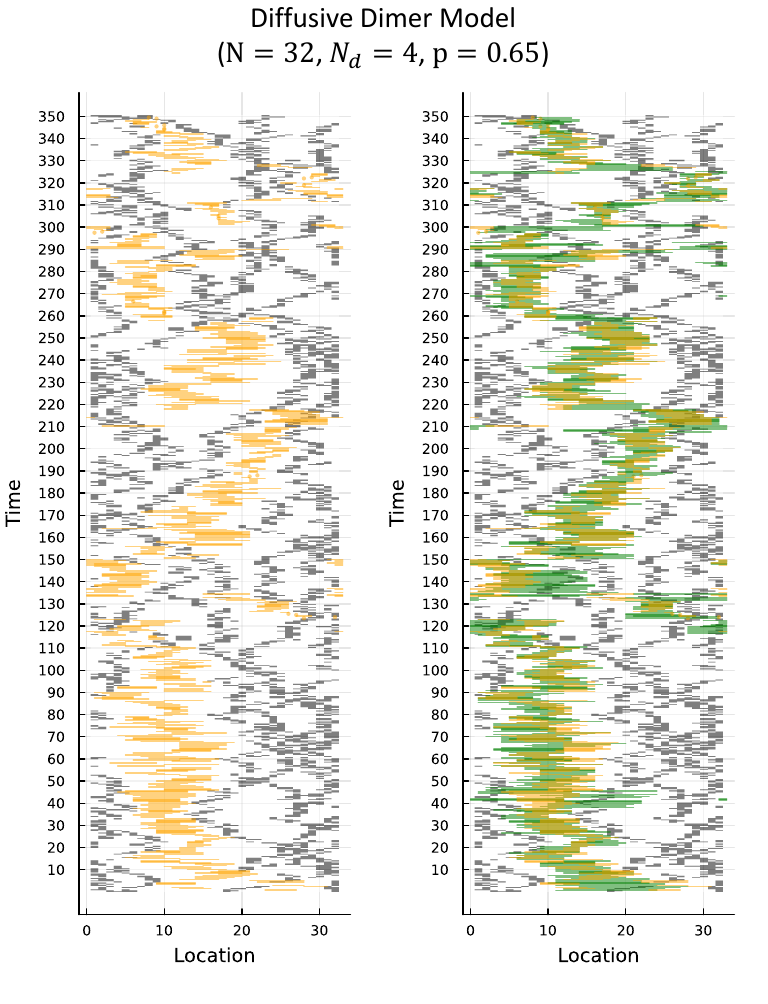}
    \caption{The shortest logical operator (left) and both the longest and the shortest logical pair (right) in the dimer model.}
  \label{fig_logical_dimer}
\end{figure}

To make a stark contrast in spatial measurer density to apparently visualize the rare regions, we next run the dimer model, where the measurers form a pair as a dimer, and each dimer evolves under SSEP rules. Measurer particle individually measures the qubit with probability $p$. We picked the number of dimers to be $N_d = N/8$, so the total number of particles is $N/4$. In this model, the AMI value peaked at $p=0.65$; hence, we picked $p=0.65$ in  Fig.~\ref{fig_logical_dimer}.

